\documentclass{svjour3}
\usepackage{amssymb}
\usepackage{epsfig}

\usepackage{natbib}

\bibpunct{(}{)}{;}{a}{}{,}

\newcommand{\be}{\begin{equation}}
\newcommand{\ee}{\end{equation}}
\newcommand{\beq}{\begin{eqnarray}}
\newcommand{\eeq}{\end{eqnarray}}

\newcommand{\ion}[2]{#1\,{\sc{#2}}}

\journalname{SSRv}

\begin{document}

\title{Future instrumentation for the study of the Warm-Hot Intergalactic Medium}

\author{Frits~Paerels \and
            Jelle~Kaastra \and
            Takaya~Ohashi \and    
            Philipp~Richter \and
            Andrei~Bykov \and
            Jukka~Nevalainen
        }

\authorrunning{Paerels et al.}
\titlerunning{Future instrumentation}

\institute{Frits Paerels 
\at Department of Astronomy and Columbia Astrophysics Laboratory,
Columbia University, 550 West 120th Street, New York, NY 10027, USA \\
                             \email{frits@astro.columbia.edu}
\and
 Jelle Kaastra \at 
 SRON Netherlands Institute for Space Research,
Sorbonnelaan 2, 3584 CA Utrecht, the Netherlands \\
 Sterrenkundig Instituut, Universiteit Utrecht
 P.O. Box 80000, NL-3508 TA Utrecht, the Netherlands
 \and
 Takaya Ohashi \at 
 Department of Physics, School of Science, Tokyo Metropolitan University,
			 1-1 Minami-Osawa, Hachioji, Tokyo 192-0397, Japan
\and
 Philipp Richter \at 
 Institut f\"uer Physik, Universit\"at Potsdam,
 			Am Neuen Palais 10, 14469 Potsdam, Germany
\and
 Andrei Bykov \at 
 A.F. Ioffe Institute of Physics and Technology, Russian Academy of Sciences,
 			26 Polytekhnicheskaya, St. Petersburg 194021, Russia
\and
 Jukka Nevalainen \at 
 Observatory, University of Helsinki, P.O. Box 14, 
 00014 T\"ahtitornim\"aki, Finland
 }

\date{Received: 19 November 2007 ; Accepted: 10 December 2007}

\maketitle

\begin{abstract}

We briefly review capabilities and requirements for future instrumentation in
UV- and X-ray astronomy that can contribute to advancing our understanding of
the diffuse, highly ionised intergalactic medium.

\end{abstract}

\section{Introduction}
\label{Introduction} 

A convergence of recent theoretical and observational work has generated rapidly
growing interest in the physics of a highly ionised intergalactic medium (IGM).
At low redshifts, this phase of the IGM likely holds the balance of the baryon
mass density not accounted for by the mass densities in stars, diffuse gas in
galaxies, the local Ly$\alpha$ forest, the Intracluster Medium (ICM) in galaxy
clusters and groups, etc. \citep{FP04}. Model calculations suggest that in fact
the baryon density in the IGM could indeed be comparable to the summed baryon
densities of the known mass components, as would be required to reach the baryon
density determined using independent arguments (big bang nucleosynthesis and the
measured abundances of the light elements, and the measured fluctuation spectrum
of the cosmic microwave background).

It is clear that the study of non-equilibrium phenomena in the low-redshift
diffuse intergalactic medium, both within and outside bound structures, can
produce a wealth of information on a large range of important astrophysical
problems, ranging from the formation of galaxies to the generation of the first
dynamically important magnetic fields. In this paper, we will attempt to
summarise the requirements and prospects for significant progress on the
observational study of these phenomena, considered in the light of planned or
proposed instrumentation. We begin with a slightly abstract discussion of
required and desired instrumental capabilities (spectral resolution,
sensitivity, etc.), before we consider specific future or proposed instruments. 

\section{X-ray and UV spectroscopy of the ICM and IGM}

\subsection{Imaging spectroscopy (emission line imaging)}

Imaging X-ray spectroscopy will play a crucial role in the study of
non-equilibrium phenomena in highly ionised diffuse media. High spectroscopic
resolution is a requirement here, not only for the measurement (if possible) of
velocity fields and the application of standard emission line plasma
diagnostics, but also simply as a means to separate extragalactic emission from
the bright and rich foreground emission from hot gas in and around our Milky Way
Galaxy, the (time variable) contribution from geocoronal and heliospheric charge
exchange excitation by the Solar wind \citep{W04} and to suppress the continuum
background from unresolved point sources and non-X-ray instrumental
backgrounds. 

To start with the latter point (suppression of continuum backgrounds for the
detection of line emission from highly ionised species in the WHIM), we need a
rough estimate for the expected emission line intensity; the extragalactic
background intensity in the soft X-ray band (below 653 eV, corresponding to
\ion{O}{viii} Ly$\alpha$ at $z=0$) can be inferred from an extrapolation of the
smooth isotropic background above 1 keV. We assume that the emission is
dominated by collisional excitation, ignoring the recombination-excited
contribution, as well as the contribution from resonant excitation by the X-ray
background continuum \citep{Ch01}. Even though the oxygen ionisation balance is
probably largely determined by X-ray photoionisation, the line emission rate
will respond strongly to heating of the gas, as it passes through the structure
formation shocks. As far as oxygen is concerned, the WHIM is a hybrid medium,
with the ionisation provided by photons, but the cooling (line emission)
provided by electron collisions. 

A simple estimate for the expected intensity in the resonance lines of
\ion{O}{vii} and \ion{O}{viii} is thus
\begin{eqnarray}
\left<I\right> & \sim & {{1}\over{4\pi}}\left<n_{\rm e}^2\right>l\,
A\, f_i \, S_{ij}(T_{\rm e}) = \\
& = & {{1}\over{4\pi}} C \left<n_{\rm e}\right>^2 l\,
A\, f_i \, S_{ij}(T_{\rm e})\ \ {\rm photons\,cm^{-2}\,s^{-1}\,sr^{-1}},
\end{eqnarray}
with $A$ the absolute elemental abundance, $f_i$ the fractions of atoms in the
relevant ionisation state, $l$ the path length, $C$ the clumping factor ($C
\equiv \left< n^2 \right> / \left< n \right>^2$). From explicit simulations (see
for instance \citealt{Da01}), we estimate the clumping factor $C \sim 100$. The
average should really be carried out explicitly over redshift as well, since the
WHIM rapidly evolves at small redshift. 

$S_{ij}(T_{\rm e}) \equiv \left<\sigma(v) v\right>$ is the thermal average of
the collisional excitation cross section $\sigma(v)$ times the electron-ion
velocity $v$. A useful discussion of thermal collisional excitation rates is
given by Mewe (\citealt{M99}; his Sect.~8.1; see also \citealt{K08} - Chapter 9,
this volume). {\sl Very} roughly, for a H-like ion, the absolute collision cross
section is of order $\pi a_0^2/Z^2$ ($a_0$ is the Bohr radius, $Z$ the nuclear
charge). For electron temperatures small compared to the excitation energy
(which is what we are dealing with here), a strong temperature dependence in
$\left<\sigma(v) v\right>$ results from the small, strongly
temperature-dependent fraction of the electrons that have sufficient energy to
excite the ions. The steep functional dependence of $S_{ij}(T_{\rm e})$ is
almost entirely determined by the temperature dependence of this fraction.
\citet{M99} gives very useful numerical expressions for the excitation rates, as
well as references to more detailed, precise data. We use these precise rates in
our estimate for $\left< I \right>$, since the analytical approximations, while
nicely reproducing the functional shape of the temperature dependence, fail in
absolute value by more than an order of magnitude for electron temperatures far
below threshold.

For the \ion{O}{viii}  Ly$\alpha$ and \ion{O}{vii} resonance lines, a very rough
estimate for  the rate is plotted as a function of electron temperature in
Fig.~1. We have assumed one-tenth Solar abundance, $f_i = 0.5$ for both ions (at
the low densities characteristic of the average WHIM, the ionisation fractions
have to be calculated from the photoionisation balance and depend mostly on
density rather than on electron temperature; strictly speaking, $f_i$ should
have been taken inside the average for the density), and $l = {1\over 3}  {\rm
c}/{\rm H}_0$, with ${\rm H}_0$ the Hubble constant, as suggested by the
simulations (the WHIM does not rise until after redshifts $z \sim 1$
\citep{Da01}, and the integral will of course be weighted towards the higher
densities, at the smaller redshifts).

\begin{figure}
\centerline{\psfig{figure=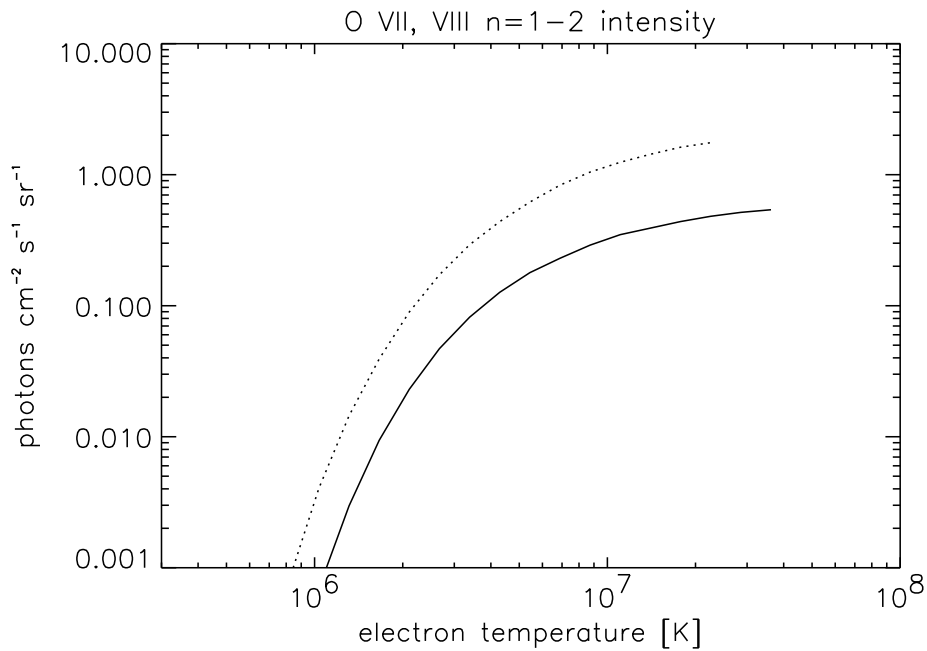,width=\textwidth}}
\caption[]{
Average photon count rates from the WHIM
in the \ion{O}{vii} $n=1-2$ triplet (dotted line) and \ion{O}{viii} Ly$\alpha$ (solid line)
lines, as a function of electron temperature. See text for explanation.
}
\end{figure}

This faint line emission has to be detected against at least the  extragalactic
continuum background (assuming it is largely unresolved, and that the background
is dominated by the astrophysical background). If we adopt an average background
continuum intensity of $\approx 40$ photons cm$^{-2}$ s$^{-1}$ sr$^{-1}$
keV$^{-1}$ at $E \sim 0.5$ keV (see \citealt{MS90}, their Fig.~12), we need an
energy resolution of order $\Delta E \lesssim 2-3$ eV at 0.5 keV in order to keep
the continuum background in a spectral resolution element comparable to the
emission line count per resolution element. Obviously, higher contrast in
emission line images requires higher spectral resolution. 

The soft background ($E \lesssim 1$ keV) is largely dominated by the foreground
emission from our Galaxy and its immediate environment. As was dramatically
demonstrated by the Wisconsin/Goddard X-ray Quantum Calorimeter (XQC) rocket
experiment \citep{M02}, most of this 'foreground' emission is in emission lines,
dominated by the \ion{O}{vii} and \ion{O}{viii} $n=1-2$ resonance lines. But
normalising simple models for the emission from $T_{\rm e} \sim 10^6$~K, Solar
abundance gas to the measured emission line intensities indicates that there
will be numerous weaker foreground lines, between which the weak redshifted,
extragalactic emission lines will have to be detected. This will likely be
impossible with data that does not at least have a resolution of $\Delta E \lesssim
2$ eV. For an impression of what the foreground spectrum might look like, see
Fig.~13 in \citet{K08} - Chapter 9, this volume. Note that due to the complexity
of this foreground, convincing detection and interpretation of the extragalactic
line emission will likely depend on a sophisticated simultaneous analysis of the
Galactic emission. 

As far as spectroscopic diagnostics are concerned, at $\Delta E \lesssim 4-5$ eV,
the important He-like \ion{O}{vii} $n=1-2$ triplet is resolved. Resolving the
thermal width of oxygen emission lines will require a very high resolving power
${\cal R}  \gtrsim 1.3 \times 10^4 (T/10^6\ {\rm K})^{-1/2}$. More likely, the
velocity dispersion observed in a finite spatial volume will be dominated by
turbulence and bulk motion in the WHIM, with characteristic amplitude probably
several 100 km s$^{-1}$, which is resolved at $\Delta E \lesssim  0.2$ eV in
observations in which the individual WHIM filaments are spatially just
resolved. 

The predicted emission line intensities are very small, and an equally important
instrument characteristic is the grasp (product of field of view $\Omega$ and
effective area $A$) of the imaging spectrometer. With an intensity of $\sim 0.1$
photons cm$^{-2}$ s$^{-1}$ sr$^{-1}$, one gets less than 10 counts in $10^6$ s
of exposure if the grasp is smaller than $A \Omega \sim 10^{-4}$~cm$^2$\,sr, or
$0.3$~cm$^2$\,deg$^2$. For an effective mapping and true physical
characterisation of the WHIM (as opposed to a single detection experiment), one
probably needs at least two orders of magnitude larger grasp.

From the large scale structure/hydrodynamics simulations, the characteristic
cross section of the WHIM filaments is about 10 Mpc, which spans an angle of
$\sim 10^{-2}$ rad, or half a degree, at redshift $z = 0.3$. To purely resolve
the WHIM structure, the angular resolution requirements therefore appear to be
modest; instead, you need a large field of view of order at least a square
degree to effectively map the structure. But low angular resolution, apart from
contaminating the signal with unresolved point sources, will also introduce
ambiguity in the precise location of the emitting ions with respect to other
components of the large scale structure. At angular resolution $\Delta\theta
\gtrsim 1-2$ arcmin, one can no longer distinguish truly diffuse intergalactic
metal line emission, from emission due to small-scale galactic winds (and the
interpretation of the data would involve assuming an explicit, complete model
for the galactic feedback on the IGM). Finally, in addition to angular
resolution requirements, any proposed optical system will also have to satisfy a
contrast (or dynamic range) requirement: if too large a fraction of the photons
are reflected or scattered far from the image core (i.e. large wings on the
point source response function), bright clusters in an image will outshine the
faint WHIM emission. This may be a non-negligible problem, since the core
densities of clusters and the average densities of the WHIM differ by a factor
$\gtrsim 10-30$, which result in a surface brightness contrast of a factor of up
to 1000.

At the time of writing, a number of CCD-based imaging experiments, on the {\sl
Chandra}, {\sl XMM-Newton}, and {\sl Suzaku} observatories, are producing data.
The spectral resolution is limited to $\Delta E \gtrsim 50$ eV, which makes it
possible to discern and partially resolve \ion{C}{vi}, \ion{N}{vii},
\ion{O}{vii} and \ion{O}{viii}, and \ion{Fe}{xvii} line emission, but not
without a model for contaminating fore- and backgrounds. Much more promising are
imaging arrays of cryogenic spectrometers (microcalorimeters and Super
Conducting Tunneling Junctions, STJ), and such  instruments are part of the
baseline for the {\sl Constellation-X} and {\sl XEUS} observatories (for general
descriptions of the operational principles of these types of detector, see for
instance \citealt{En05}). In addition, dedicated soft X-ray, wide
field microcalorimeter-based experiments have been proposed with the explicit
aim of finding and characterising the WHIM ({\sl DIOS, NEW, EDGE}). A brief
description of these future instruments follows. 

Both {\sl Constellation-X}\footnote{http://constellation.gsfc.nasa.gov/} (NASA),
and {\sl XEUS}
\footnote{http://www.rssd.esa.int/index.php?project=XEUS\&page=index}  (ESA) are
designed as general facilities,  with a  wide energy band, up to 10 keV (and a
hard X-ray telescope above $10-15$ keV, on {\sl Constellation-X}), and,
consequently, a rather long focal length and relatively small field of view (10
m focal length and F/8 focal ratio for {\sl Constellation-X}; 35 m focal length
for {\sl XEUS} and $\sim$ F/10 focal ratio for {\sl XEUS}). In the {\sl
Constellation-X} focal plane, an array of microcalorimeters provides imaging
spectroscopy. In the baseline designs, the microcalorimeters are of the
Transition Edge Sensor (TES) variety, with a design energy resolution goal of 
$\Delta E = 2$ eV. The field of view is approximately 5 arcmin in diameter, and
the spectrometer has an effective area of $\approx 1$ m$^2$ at sub-keV energies,
achieved by flying four identical telescopes in parallel. {\sl Constellation-X}
also has a novel reflection grating spectrometer, which can be applied to high
resolution spectroscopy of point sources. {\sl XEUS} has been designed with two
different types of cryogenic spectrometers in mind, Superconducting Tunneling
Junctions and TES microcalorimeters; the latter are currently the baseline. The
energy resolution goals are 1 eV (STJ) and 2 eV (TES). The field of view of the
cryogenic spectrometer will be approximately 0.75 arcmin diameter, with an
effective area $\sim 5$~m$^2$ at low energies. Note that all cryogenic
spectrometers need extensive blocking of UV/optical/infrared radiation, and the
possible beam filters all have considerable low energy X-ray photoelectric
absorption, resulting in sensitivities that steeply decline with decreasing
photon energy. The effective areas listed above therefore are no more than easy
reference numbers; for detailed sensitivity calculations, be sure to consult the
relevant observatory technical information. 

From the above characteristics, it is clear that the cryogenic imaging
spectrometers on {\sl Constellation-X} and {\sl XEUS} will have sufficient
energy resolution, but that their efficiency is limited. The relevant grasps are
$\sim 50$~cm$^2$\,deg$^2$, and $\sim 6$~cm$^2$deg$^2$ for {\sl Constellation-X}
and {\sl XEUS}, respectively. Note that these figures do not include the
transmission of the optical blocking filters, which lowers the grasps by factors
of a few at the X-ray energies of interest. In addition, the fields of view are
not ideally matched to the problem. We will probably see line emission from the
WHIM in (possibly) most deep observations, but only from a series of individual
lines-of-sight. 

By contrast, there has also been a number of proposals for smaller missions,
that aim at detection of the WHIM line emission with a dedicated wide field of
view, short focal length, high spectral resolution, soft X-ray imaging
spectrometer. These characteristics (with the exception of the spectral
resolution) are interlocking: a short focal length promotes sensitivity to faint
emission against non-astrophysical background, allows a large field of view, but
limits the energy response to low X-ray energies (which is ideal for study of
the WHIM). Coupled with a high resolution imaging spectrometer, this is the
natural design for study of extended diffuse faint soft X-ray line emission. We
will therefore describe these mission concepts in somewhat greater detail. 

The {\sl Diffuse Intergalactic Oxygen Surveyor (DIOS)} concept originated in
Japan \citep{O06}. It is built around an imaging array of TES microcalorimeters,
with an energy resolution of 2 eV, a 50 arcmin diameter field of view, an
effective area in the O K band of $> 100$ cm$^2$, which gives a grasp of order
100~cm$^2$\,deg$^2$. The mission lifetime should be $> 5$ yrs. With this
grasp and field of view, the WHIM should clearly come into view; the total
fraction of the baryonic mass density probed could be as high as 20~\%. 

A similar concept was studied by the Dutch: {\sl NEW}, or the {\sl Netherlands
Explorer of the Web} \citep{denherder06}. In outline, it is very similar to {\sl
DIOS}. It has a 24$^2$ array of TES microcalorimeters, viewing a $1 \times 1$
deg$^2$ field of view at 2 eV energy resolution. The effective area is
substantially larger than {\sl DIOS}, with a 500 cm$^2$ baseline design in the O
K band.

Other proposed missions incorporate the imaging spectroscopy of the
high-ionisation metal emission lines with arrays of cryogenic spectrometers, but
complement this with absorption line spectroscopy ({\sl Estremo} and {\sl
EDGE}). For convenience, we will discuss these missions below, in the section on
X-ray absorption spectroscopy. 

Finally, the broad Hydrogen Ly$\alpha$ absorption lines discussed in
\citet{Richter08} - Chapter 3, this volume, trace the (small) neutral H content
of the highly ionised medium. One can search for the corresponding faint
Ly$\alpha$ emission lines, which would have to be done with a high-resolution UV
imaging spectrometer. Work is proceeding on a balloon-borne experiment, to
perform a first survey of line emission in the $2000-2200$~\AA\ band; this band
is accessible from a balloon, and is limited by strong atmospheric absorption on
either end ({\sl FIREBALL}, or {\sl F}aint {\sl I}ntergalactic {\sl R}edshifted
{\sl E}mission {\sl BALL}oon
\footnote{http://www.srl.caltech.edu/sal/igm\_project.htm}). With a proven
spectrometer concept, a survey from space could then follow up at shorter
wavelengths. The $2000-2200$~\AA\ window contains H Ly$\alpha$ $\lambda$1216
\AA\ at $z = 0.7$, \ion{C}{iv} $\lambda$1550 \AA\ at $z = 0.36$, and \ion{O}{vi}
$\lambda$1033 \AA\ at $z = 1.0$. The instrument, based on an echelle
spectrograph, images approximately $200 \times 200$ arcsec$^2$, and can send the
light of 20 individual, contiguous 7 arcsec wide strips through the
spectrograph. The resolving power set by the 7 arcsec slit size is ${\cal R} =
8000$ in the first version of the instrument. The sensitivity is such that a
diffuse Ly$\alpha$ intensity of $\sim 500$
photons\,cm$^{-2}$\,s$^{-1}$\,sr$^{-1}$ can be detected in a single night, which
corresponds to emission from regions of overdensity $\delta \sim 300-1000$.

\subsection{Absorption line spectroscopy}

\subsubsection{Ultraviolet}

As discussed by \citealt{Richter08} - Chapter 3, this volume, UV absorption
spectroscopy  of intervening hydrogen and metal absorbers currently represents
the most sensitive method to study the  statistical and physical properties of
warm-hot intergalactic gas at low redshift. During the last years, UV spectra of
several low-redshift QSOs obtained with the {\sl Far Ultraviolet  Spectroscopic
Explorer (FUSE)} and the {\sl Space Telescope Imaging Spectrograph (STIS)}
aboard the {\sl Hubble Space Telescope (HST)} have been used to study the
properties of WHIM absorbers in great detail. By now, however, both instruments
unfortunately are out of commission due to technical problems and therefore
additional UV observations of the WHIM have to await future space-based UV
instruments, which are discusssed in the following.

While the spectral resolution of the former {\sl FUSE} and {\sl STIS} UV
spectrographs (${\cal R}\approx 20\,000-45\,000$) were sufficient to reliably
identify and analyse absorption features from the WHIM at low redshifts (in
contrast to current X-ray spectrographs; see \citealt{Richter08} - Chapter 3,
this volume), the most important requirement for the next-generation UV
spectrographs is a substantial gain in sensitivity. Only with more sensitive UV
instruments will it be possible to achieve better statistics on the distribution
and properties of WHIM absorbers along a large number of QSO sightlines at low
and intermediate redshifts and to  obtain spectra with higher signal-to-noise
ratios. The latter aspect, in particular, is crucial to minimising systematic
errors in the data analysis and interpretation of the UV absorption signatures.

Fortunately, the next UV spectrograph most likely will become  available already
in late 2008, when the {\sl Cosmic Origins Spectrograph (COS)} will be installed
on {\sl HST} during the fifth and final service mission (SM$-$5) of {\sl HST}
(note that  SM$-$5 is nothing but the reinstated SM$-$4 that was canceled in
2004). SM$-$5 (flight STS$-$125) currently is scheduled for August 7, 2008 (see
{www.nasa.gov} for regular updates on the NASA space shuttle launch schedule).
{\sl COS} is designed for high throughput, medium-resolution (${\cal R} \sim
20\,000$) spectroscopy of point sources in the UV wavelength range. The {\sl
COS} instrument has two channels, a far-UV channel covering $1150-1775$ \AA, and
a  near-UV channel operating in the range $1750-3000$ \AA.  The {\sl COS}
instrument has been built with maximum effective  area as primary constraint. It
will provide a gain in sensitivity of more than one order of magnitude compared
to {\sl STIS} and other previous spectrographs (e.g., {\sl GHRS}) installed on
the {\sl HST}.  {\sl COS} is therefore expected to deliver a large number of 
high-quality spectra of low- and intermediate-redshift QSOs and AGN. These data
will allow us to better understand the baryon budget  and physical conditions in
intervening \ion{O}{vi} systems and broad Ly\,$\alpha$ absorbers (BLAs) and the
space distribution  of warm-hot intergalactic gas in relation to galactic
structures. Note that it is also planned to repair {\sl STIS} during the same service 
mission, so that there will be two powerful UV spectrographs with different
spectral capabilities available at the {\sl HST} at the same time.

Next to the {\sl COS} and {\sl STIS} instruments aboard the {\sl HST}, there
possibly will be another powerful UV instrument available at the beginning of
the next decade. The Russian-led  {\sl World Space Observatory (WSO)} is an 
independent UV space telescope and is designed to represent  an international
follow-up of the {\sl HST} mission. Next to Russia, other countries such as
China, Spain, Italy, Germany are also involved in the {\sl WSO} project and
partly contribute with either money and/or instrumentation. The {\sl WSO} will
be equipped with a 1.7~m primary mirror, two high-resolution spectrographs with
${\cal R} \sim 50\,000-55\,000$ operating in the FUV and UV wavelength bands
$1020-1720$ \AA, and $1740-3100$ \AA, and a low-resolution spectrograph (${\cal
R} \sim 2\,500$) for the range $1020-3100$ \AA\, (for details and further
information see {\sl WSO} homepage.\footnote{http://wso.inasan.ru}  A big
advantage of the {\sl WSO} is that it represents an instrument entirely
dedicated to the UV range. This will enable observers also to aim for long
integration times of relatively faint (i.e. distant) UV background  sources
without facing competition with observers in other wavelength bands (as is the
case for {\sl COS} on {\sl HST}, for instance). In addition, {\sl WSO}
observations of intermediate-redshift QSOs will be particularly important to
search for moderately redshifted WHIM EUV absorption from \ion{Ne}{viii} and
\ion{Mg}{x} (see \citealt{Richter08} - Chapter 3, this volume, Sect.\,2.2.1)
that can be observed in the {\sl WSO} FUV band. Consequently, the {\sl WSO} will
represent another very powerful UV instrument to study ultraviolet absorption
signatures of the  WHIM at low and intermediate redshift. 

Next to {\sl STIS}, {\sl COS}, and the {\sl WSO}, which by now represent
approved science missions, there are various future mission concepts for
space-based UV instruments, for instance as part of NASA's Origins
Program.\footnote{ 
http://www.nasa.gov/home/hqnews/2004/jul/HQ-04246\_mission\_concepts.html} Among
these concepts, the {\sl Baryonic Structure Probe} (PI: Kenneth Sembach, STScI)
and the {\sl High Orbit Ultraviolet-visible Satellite (HORUS}; PI: Jon Morse,
Arizona State University) represent particularly interesting concepts for future
UV observatories, as these instruments would be able to investigate warm-hot
intergalactic gas  in the ultraviolet in {\sl both} absorption and emission.

\subsubsection{X-ray}

As discussed in \citealt{Richter08} - Chapter 3, this volume, detecting X-ray
line absorption from the highly ionised light- and medium-$Z$ elements in the
WHIM (in practice, C, perhaps N, O, Ne, perhaps Fe) is challenging, due to the
fact that the expected equivalent widths are very small, as well as to the fact
that the absorption is expected to be sparse, at any spectrometer sensitivity
currently foreseen. The absorption will be sparse in two senses: a single
absorbing filament of WHIM gas will produce only one, or at most a few
absorption lines; and on any line of sight to a bright, distant background
continuum source we expect to be able to see only of order a few filaments at
most. It is conceptually somewhat misleading to refer to the intergalactic X-ray
absorption lines as the 'X-ray Forest' counterpart to the Ly$\alpha$ Forest
\citep{H98}. Nevertheless, X-ray absorption spectroscopy of the WHIM complements
study of its line emission in very important ways: the sensitivity of absorption
spectroscopy, for a given spectrometer sensitivity, only depends on the
continuum flux of the background source, and, given sufficiently bright sources,
studying the WHIM in absorption is potentially more sensitive to small column
density filaments, as well as more distant filaments. Detecting the line {\sl
emission} from distant and/or low-density filaments will always be difficult.
Finally, it is possible in principle to build (very) high resolution slitless
grating spectrometers for point-source spectroscopy, and one could study the
velocity structure (and hence the physical structure) of individual filaments,
and the relation between, for instance, Ly$\alpha$ and \ion{O}{vi} absorbers,
and more highly ionised X-ray absorbers. Performing similarly high resolution
diffuse emission line spectroscopy will probably be even much harder. On the
other hand, given the sparseness of suitable backlighter sources, the absorption
spectroscopy will most likely not reveal much about the full 3D topology of the
WHIM, as the emission line imaging naturally does. 

Detecting the weak X-ray absorption lines will require high spectral resolution
and high sensitivity, the latter to ensure a sufficiently large number of
photons per resolution element, as well as to make sure sufficiently weak (and
therefore numerous) sources can be used, thereby covering both the sky as well
as providing enough redshift depth.  The appropriate scale for the spectral
resolution is set by the thermal width of the lines. In the worst case (no large
scale or turbulent velocity fields) the C, N, O, and Fe resonance lines will
rapidly saturate, and the equivalent width will approximately stall at the value
it attains at the thermal width. The thermal width, expressed as the Gaussian
velocity dispersion, is  $\sigma_v = \sqrt{kT/m_i} = 22 (T/10^6\ {\rm K})^{1/2}$
km s$^{-1}$ for oxygen ions. In  \citealt{K08} - Chapter 9, this volume (their
Fig.~9), we show the curve of growth for \ion{O}{vii} and \ion{O}{viii}
resonance lines; at the thermal width, these first saturate at equivalent width
$W_{\lambda} \approx 4$ m\AA, which corresponds to a resolving power of 5000.
The equivalent number for the resolution in energy measure is $\Delta E \approx
0.1$ eV. 

In \citealt{Richter08} - Chapter 3, this volume, we reproduced a cumulative
probability distribution, derived from large scale structure/hydrodynamics
simulations, for the \ion{O}{vii} and \ion{O}{viii} column densities. At
equivalent width $W_{\lambda} = 4$ m\AA, the expectation value for the number of
filaments per unit redshift producing that EW or larger on a random line of
sight is approximately $dN/dz \approx 3$. A requirement for a future experiment
could therefore be phrased as 'the capability to detect $\lesssim 4$ m\AA\ 
equivalent width absorption lines in multiple high redshift ($ z \gtrsim 1$)
continuum sources'. 

The grating spectrometers on {\sl Chandra} and {\sl XMM-Newton} do not have
sufficient resolution; they have $\Delta\lambda \approx 50$ m\AA\ at 20 \AA\
(the {\sl Chandra} HETGS has better resolution, but only very limited efficiency
in the O K band). The results of searches on very bright targets with these
instruments are described in \citet{Richter08} - Chapter 3, this volume. 

The imaging X-ray spectrometers discussed above can also be used for absorption
spectroscopy, of course. The relatively coarse energy resolution ($\Delta E \sim
1-$few eV) is not better than the resolving power we currently have with the
grating spectrometers on {\sl Chandra} and {\sl XMM-Newton} ($\Delta\lambda =
0.05~$\AA\ corresponds to $\Delta E = 1.5$ eV at 20~\AA), and the improvement in
sensitivity would come entirely from the larger effective area. In order to
significantly improve our understanding of the WHIM by X-ray absorption
spectroscopy with practicable cryogenic spectrometers, the effective area should
therefore be appreciably larger than those of the {\sl Chandra} and {\sl
XMM-Newton} grating spectrometers. Nevertheless, a survey based on moderate
sensitivity spectroscopy of, roughly speaking, the extragalactic sources in the
4U catalog is not likely to produce much relevant information, beyond bare
detection of the WHIM. 

Even though the energy resolution of the microcalorimeter spectrometer
envisioned for {\sl XEUS} is not ideal for the intergalactic absorption
spectroscopy, pointing 5 m$^2$ of spectroscopic effective area at a bright
continuum source will of course produce a very high signal-to-noise continuum,
in which absorption lines of equivalent width well below the formal
spectroscopic resolution will be detectable in principle. The recent attempts
with the grating spectrometers on {\sl Chandra} and {\sl XMM-Newton} under
similar circumstances have illustrated how important a very detailed
understanding of the continuum sensitivity of the spectrometer is; in fact, in
practice this understanding could well be the ultimate limiting factor on the
sensitivity, not the photon counting fluctuations or background characteristics.
The goals for intergalactic absorption spectroscopy with {\sl XEUS} have been
summarised by \citet{Bar03}: a spectrum on each of about 50 AGN at redshifts $z
\gtrsim$ few tenths, sensitive enough to detect oxygen resonance absorption
lines of equivalent width $\sim 0.4$ eV.\footnote{for a recent update, see the
{\sl XEUS} Science Requirements Document, 4th edition,
http://www.rssd.esa.int/index.php?project=XEUS\&page=Multimedia, p.~11}

Using the microcalorimeter spectrometer on {\sl Constellation-X} for absorption
spectroscopy faces a similar problem, since the projected energy resolution is
very similar to that for {\sl XEUS}. But the baseline design for the observatory
includes a reflection grating spectrometer with substantially higher spectral
resolution (though smaller effective area; \citealt{Cot06}). The baseline design
calls for a resolving power of at least 300 at 20 \AA, with a goal of 3000. The
effective area should exceed 1000 cm$^2$, so that the sensitivity should be at
least a factor 10 higher than currently reachable with the Reflection Grating
Spectrometer on {\sl XMM-Newton}. This is illustrated graphically on the {\sl
Constellation-X}
website\footnote{http://constellationx.nasa.gov/science/missing\_baryons/whim.html};
but note that the prominent absorption line in that figure (which presumably
displays the capability of the {\sl Constellation-X} grating spectrometer) is
the \ion{O}{vii} $n=1-2$ resonance line at redshift $z=0$ (from gas in a
putative hot halo of our Galaxy and/or the intragroup medium of the Local
group), which has an equivalent width of $\approx 13$ m\AA\
(\citealt{Williams05,Ras07}), a factor 3 larger than the strongest
true intergalactic absorption lines we expect.

\begin{figure}
\centerline{\psfig{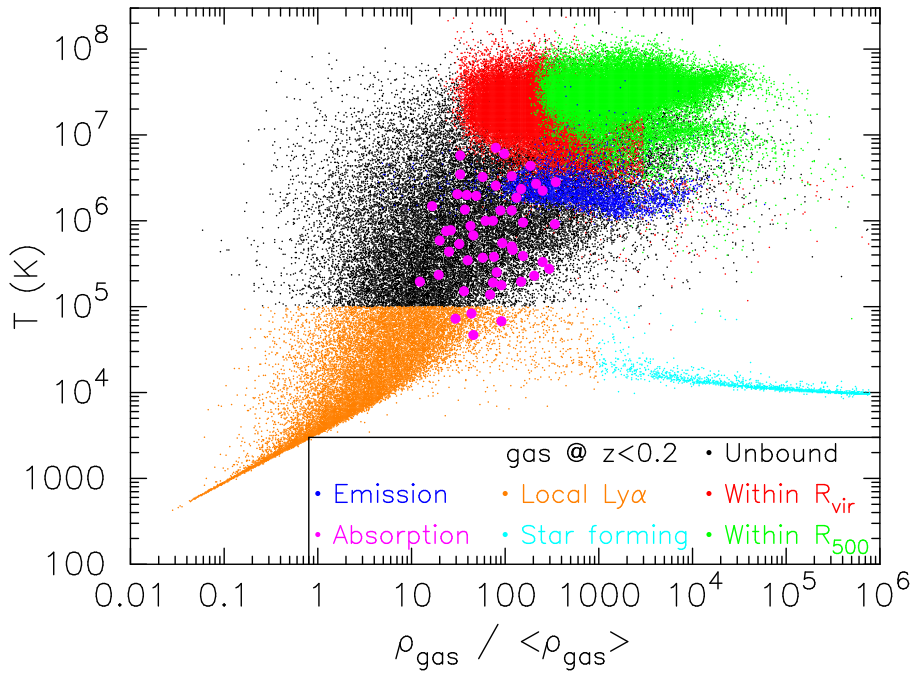}}
\caption{ Phase diagram of the intergalactic medium at $z < 0.2$. The dots are
values of electron temperature and overdensity recorded in the computational
volume of a large scale hydrodynamics simulation for $ z < 0.2$; the various
colors indicate different characteristic components of the diffuse gas: the
warm, photoionised Ly$\alpha$ forest, the gas in virialised structures (groups
and clusters), and the cool gas in (star-forming) galaxies. The WHIM-like points
are black and dark blue. The dark blue points indicate the region of the
density/temperature plane that will be probed in emission by the cryogenic
imaging spectrometer on {\sl EDGE}; the purple dots indicate the region that can
be probed in absorption towards bright GRB afterglows in a 3 year core mission.
From \protect\citet{P07}. 
}
\end{figure}

The problem of a sufficiently large population of suitably bright point sources
at non-trivial redshifts can be solved by using a 'renewable' set of sources:
gamma-ray burst afterglows. If one catches the afterglow early enough, the
cumulative fluence over the course of the burst decay compares favorably with
the fluence of the brightest extragalactic steady sources, observed over a
similar time interval. There is no limit to the number of sources this way;
moreover, GRB happen to originate on average at fairly high redshift ($z \gtrsim
1$), which is ideal for the WHIM problem. But in order to take advantage of
these characteristics, one needs to be able to point at a GRB afterglow
literally within a matter of minutes. The {\sl
Pharos}\footnote{http://hea-www.harvard.edu/$\sim$elvis/Pharos\_GSFC\_Aug06.pdf};
\citep{EF03} and {\sl Estremo} \citep{P06} mission concepts build on the
all-sky GRB monitoring and rapid repointing scheme first employed by {\sl
Swift}. {\sl Pharos} employs a relatively short focal length, moderate angular
resolution soft X-ray telescope to feed high reflectivity, high dispersion
reflection gratings, of the so-called off-plane variety (see for instance
\citealt{McE04}), and reading out in the third spectral order. The spectrometer
is designed to reach the thermal width of the oxygen absorption lines ($\sim 4$
m\AA\ at 20~\AA). {\sl Estremo}, instead, relies on non-diffractive spectroscopy,
with an array of TES microcalorimeters. 

More recently, the {\sl Estremo}, {\sl NEW}, and {\sl DIOS} ideas have
tentatively been merged into a new proposal to ESA's {\sl Cosmic Vision}
program, under the name {\sl EDGE} (\citealt{P07}; {\sl E}xplorer of {\sl
D}iffuse emission and {\sl G}amma-ray burst {\sl E}xplosions). The {\sl EDGE}
concept relies on rapid repointing to perform absorption spectroscopy on
gamma-ray burst afterglows, with a TES-microcalorimeter-based spectrometer, with
baseline energy resolution $\Delta E = 3 $ eV. The same array of
microcalorimeters views an area of $0.7 \times 0.7$~deg$^2$ of the sky, with an
effective area of 1000~cm$^2$, which translates to a grasp of
$500$~cm$^2$\,deg$^2$. With this instrument, the observatory will carry out an
imaging soft X-ray emission line survey as part of its core mission; part of the
available exposure time will be dedicated to imaging the WHIM on lines of sight
for which there is a significant absorption spectrum. There is compelling
scientific rationale for combining absorption and emission spectroscopy. If
nothing else, determining both the optical depth and the emission measure in a
given transition gives a rough measurement of linear size and ion density of the
system; with assumptions on the atomic abundance, one can derive the total
density. This, the largest and most ambitious of the various proposed WHIM
experiments, will probe\footnote{or rather, could have probed; ESA did not
select the concept for the Cosmic Vision program in the Fall of 2007} a
significant fraction of the phase plane of the low-density IGM at $z \lesssim 1$
(see Fig.~2). 

\subsection{Comparisons}

We will attempt to summarise the capabilities of the specific X-ray
instrumentation discussed above in a single graph. The sensitivity to detection
of either weak emission or absorption lines will roughly scale as
$(A\Omega/\Delta E)^{1/2}$, with $A$ and $\Omega$ the effective area and field
of view, and $\Delta E$ the spectral resolution, for the purely counting
statistics-dominated case. For each of the instruments, we plot this figure of
merit against the spectral resolution in Fig.~3. We argued that the requirements
for detecting WHIM emission or absorption limit the viable experiments to the
region $\Delta E < 2$ eV, $A\Omega > 30$ cm$^2$\,deg$^2$, or $(A\Omega/\Delta
E)^{1/2} > 2.3$ m$\,$arcmin\,eV$^{-{1/2}}$. That shows that the future large
observatories {\sl XEUS} and {\sl Constellation-X} are right at the lower end of
the requirements; a large dedicated experiment such as {\sl EDGE} has the
required capability.

\begin{figure}
\centerline{\psfig{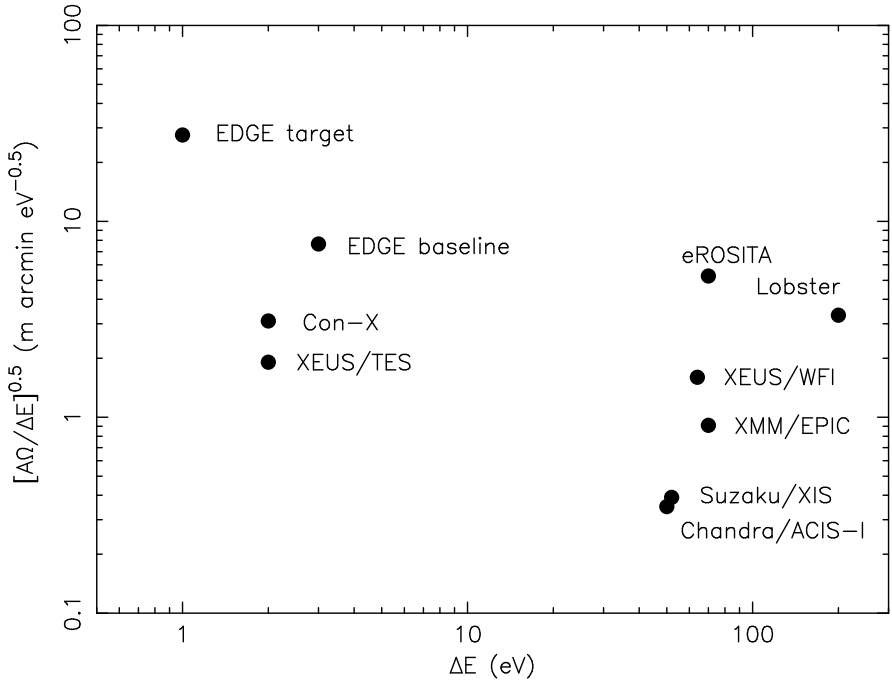}}
\caption{
A figure of merit for the detection of faint line absorption or emission from
the WHIM, in the statistics-limited regime, the square root of the grasp divided
$A\Omega$ by the spectral resolution $\Delta E$, in units
(m$^2$\,arcmin$^2$\,eV$^{-1}$)$^{0.5}$, versus spectral resolution $\Delta E$,
for a number of planned and currently operating X-ray imaging spectroscopic
observatories. 
}
\end{figure}

\subsection{Spectroscopy: a warning}

It has been emphasised throughout the papers in this volume that advancing our
understanding of the low-density IGM (outskirts of clusters and the WHIM) means:
determining the physical properties (density, temperature, abundances, possibly
velocity fields) of the medium to an accuracy sufficient to uniquely trace total
mass and thermodynamical/enrichment history. A mere detection, and a rough
determination of the total mass density in the WHIM, is not sufficient
motivation to warrant the (potentially large) expenditure of funds and other
resources on the experiments required to address this problem. Also, a proper
study of the astrophysics of the low-density, highly ionised IGM naturally
complements studies in other fields (galaxy formation and evolution, signatures
of cosmic feedback from star formation and supermassive Black Holes, maybe even
reionisation) and therefore forms a natural component of modern astrophysics.

\begin{figure}    
\begin{center}
\hbox{
\psfig{file=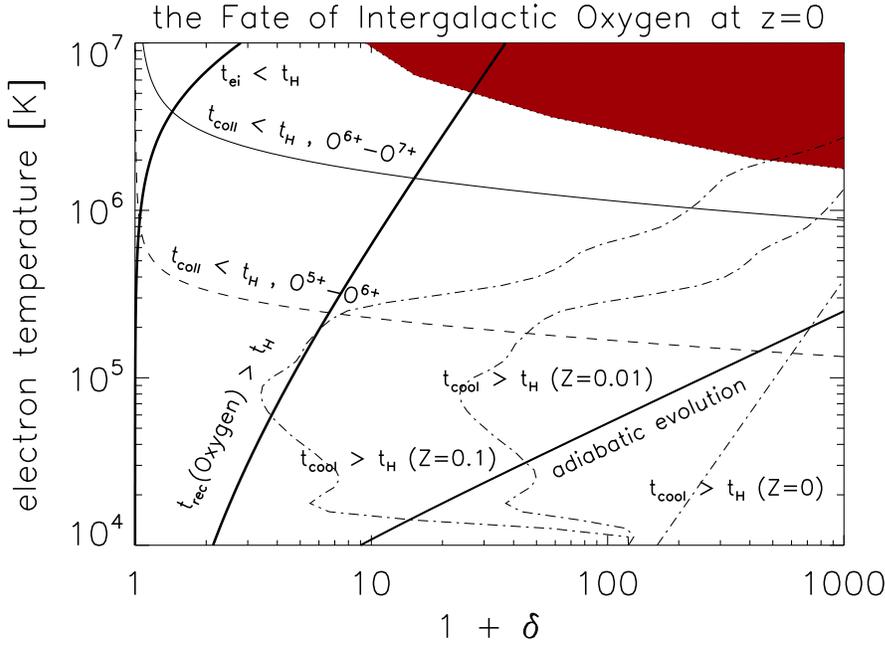,width=\textwidth,clip=}}
\caption{Phase diagram for oxygen in the IGM at redshift $z=0$. Density is
parameterised as $\delta =  \rho/\left< \rho \right>$. The various boundaries
separate regimes under which a certain process does, or does not attain
equilibrium over a Hubble time. Critical boundaries for kinetic and thermal
equilibrium are: the solid line in the upper left hand corner labeled '$t_{\rm
ei} < t_{\rm H}$' indicates where electron and proton fluids reach kinetic
equilibrium (proton temperature equal to electron temperature) in a Hubble time
($t_{\rm H}$): at low density and high temperature, such equilibrium does not
obtain. Low density gas will not radiatively cool over a Hubble time to the left
of the boundaries marked '$t_{\rm cool} >  t_{\rm H}$'; the three curves are
labeled with the metallicity, $Z$, expressed as a fraction of Solar metallicity.
The cooling time was calculated for collisionally ionised gas. The solid curve
labeled '{\sl adiabatic evolution'} indicates the locus of gas that has only
undergone adiabatic compression or expansion since high redshift (initial
condition $T_{\rm e} \sim 10^4$ K); all shock-heated gas will be above this line
right after passing through a shock. Critical boundaries for the ionisation
equilibrium of oxygen are: the shaded area in the upper right hand corner
indicates the regime where the collisional ionisation timescale is shorter than
the photoionisation timescale, for ionisation
\ion{O}{viii}$\rightarrow$\ion{O}{ix} (O$^{7+}\rightarrow$O$^{8+}$). The two
boundaries labeled '$t_{\rm coll}  < t_{\rm H}$' indicate where the collisional
ionisation timescale becomes shorter than the Hubble time. At lower temperature,
the ionisation balance cannot be in (collisional) equilibrium. Upper (solid)
curve is for ionisation \ion{O}{vii}$\rightarrow$\ion{O}{viii}
(O$^{6+}\rightarrow$O$^{7+}$), lower (dashed) curve for 
\ion{O}{vi}$\rightarrow$\ion{O}{vii} (O$^{5+}\rightarrow $O$^{6+}$). The steep
solid curve labeled '$t_{\rm rec}{\rm (Oxygen)} > t_{\rm H}$' indicates where
the radiative recombination timescale (\ion{O}{ix}$\rightarrow$ \ion{O}{viii} or
O$^{8+} \rightarrow$O$^{7+}$) exceeds the Hubble time (no recombination at low
densities). }
\label{}
\end{center}  
\end{figure}

We therefore emphasise again the importance of high resolution spectroscopy,
spatial resolution, and sky coverage. Below, we reproduce the phase diagram of
metals in the low-density IGM at $z \sim 0$, as shown in \citet{By08}, to
emphasise the fact that measurements and conclusions based on simple assumptions
concerning the physical state of the metals in the highly ionised IGM are likely
to yield biased and unreliable results. For instance, even detection of both an
\ion{O}{vii} $n=1-2$ resonance absorption line as well as a corresponding
\ion{O}{viii} Ly$\alpha$ absorption line from a single filament, does not
uniquely translate into an electron temperature, unless one knows what the
dominant ionisation mechanism is. And in fact, it is quite possible that the
ionisation balance between H- and He-like oxygen has not reached equilibrium in
the filament under study, if its density is very low. Likewise, at limited
spatial resolution, we may confuse the topology of line emitting and absorbing
gas, relative to the distribution of galaxies, and our images are confused by
metals freshly injected into the IGM by actively star-forming galaxies. 
Finally, at limited sensitivity, we will collect a sample of absorption line
measurements and emission line images that will necessarily be biased towards
the most massive filaments (the emission line images even more so than the
absorption line detections), as well as the most nearby ones (in emission).
Given the very large spread in density, temperature, and enrichment expected in
the WHIM, we should keep in mind that sampling the properties of the most
massive, densest phases of the medium is highly biased, and our measurements
should reach well below these 'tips of the iceberg' before we can claim to have
characterised the WHIM.

\section{Hard X-ray and $\gamma$-ray emission}

We conclude with a brief, non-exhaustive description of future instrumentation
for the detection of hard X-ray and $\gamma$-ray emission associated with
nonthermal processes, as described in \citealt{Petrosian08} - Chapter 10, this
volume. The potential of {\sl GLAST} for detection of high energy $\gamma$-rays
is briefly addressed in this latter reference; here, we concentrate on detection
of hard X-ray emission (traditionally approximately the $10-100$~keV band). The
challenge is to detect faint, extended diffuse emission over a large ($\lesssim$
several arcminutes) field of view in the presence of strong detector
backgrounds. Traditionally, instruments in this band have operated with
collimators or coded masks, because it is difficult to focus hard X-rays for
true imaging. However, the coming generation of hard X-ray experiments is built
around innovative focusing optics, which promises much higher contrasts for
astrophysical sources, and the detection of hard nonthermal emission from
clusters is one of the primary science areas for these new instruments.

{\sl Simbol-X} is a hard X-ray mission, operating in the $0.5-80$~keV range,
proposed as a collaboration between the French and Italian space agencies with
participation of German laboratories for a launch in 2013 \citep{Ferrando06}.
Relying on two spacecraft in a formation flying configuration, {\sl Simbol-X}
uses a $20-30$~m focal length X-ray mirror to focus X-rays with energies above
10 keV, resulting in over two orders of magnitude improvement in angular
resolution and sensitivity in the hard X-ray range with respect to non-focusing
techniques. The field of view of the instrument (at 30 keV) is $> 12'$ in
diameter. The on-axis sensitivity is better than
10$^{-14}$~erg\,cm$^{-2}$\,s$^{-1}$ (i.e. 0.5 $\mu$Crab) in the $10-40$ keV
band, providing a 3$\sigma$ detection in 1 Ms for a power law source spectrum
with a photon index of 1.6. The goals for on-axis effective area S are: at 0.5
keV photon energy  S $>$ 100  cm$^2$, at 2 keV  S $>$ 1000 cm$^2$, at 8 keV S
$>$ 600 cm$^2$,   at 30 keV S $>$ 300 cm$^2$, at 70 keV S  $>$ 100 cm$^2$,  and
at 80 keV S $>$ 50 cm$^2$.

The focal plane has two detector systems: the Low Energy Detector (LED) and the
High Energy Detector (HED). The heart of the low energy detector is a monolithic
macro pixel detector array of 128 $\times$ 128 pixels. The readout scheme of the
pixel detector is that of an active pixel sensor. An active pixel sensor
consists of a Silicon drift detector with an integrated DEPFET for readout. An
optical blocking filter is deposited directly onto the surface of the LED. The
LED detector is made of a single Si wafer on which the 128 $\times$ 128 pixels
are integrated. The HED is a hard X-ray camera made of an array of modules of 1
cm$^2$ arranged in a 8 $\times$ 8 square to cover the field of view. The
detector material is CdTe (or CdZnTe) which provides the stopping power
necessary to detect the hard X-rays. This material has flight heritage on {\sl
INTEGRAL} and {\sl SWIFT}. The broad sensitivity calculated for the $10-40$ keV
band provides a flux limit of about 6$\times$ 10$^{-15} \rm
~erg\,cm^{-2}\,s^{-1}$ that can be reached in 1 Ms.

The {\sl NuSTAR} \citep{Harr05} and {\sl NEXT} missions \citep{Naka06} similarly
feature focusing hard X-ray optics. 

\begin{acknowledgements}

The authors express their gratitude to ISSI for its support of the team
'Non-virialized X-ray components in clusters of galaxies'. FP acknowledges
support from the Dutch Organization for Scientific Research NWO, and expresses
his gratitude to SRON, the Netherlands Institute for Space Research, for its
hospitality. 

\end{acknowledgements}

\end{document}